\RequirePackage{fix-cm}
\documentclass[smallextended]{svjour3}       
\smartqed  
\usepackage{graphicx}
\begin{document}

\title{$N^*$ Production from  $e^+e^-$ Annihilations \thanks{This work is supported in part by DFG and NSFC through funds provided to the Sino-German CRC 110 ``Symmetry and the Emergence of Structure in QCD" (NSFC Grant No. 11621131001), as well as an NSFC fund
under Grant No. ~11647601.}
}


\author{Bing-Song Zou
}


\institute{Bing-Song Zou \at
              CAS Key Laboratory of Theoretical Physics, Institute
      of Theoretical Physics, and University of Chinese Academy of Sciences, Chinese Academy of Sciences, Beijing 100190, China \\
              Tel.: +86-10-62562584\\
              Fax: +86-10-62562587\\
              \email{zoubs@itp.ac.cn}           
}

\date{Received: date / Accepted: date}

\maketitle

\begin{abstract}
Up to now, the $N^*$ production from  $e^+e^-$ annihilations has been studied only around charmonium region. Charmonium decays to $N^*$s are analogous to (time-like) EM form factors in that the charm quark annihilation provides a nearly pointlike (ggg) current. Complementary to other sources, such as $\pi N$, $eN$ and $\gamma N$ reactions, this new source for $N^*$ spectroscopy has a few advantages, such as an isospin filter and a low spin filter. The experimental results on $N^*$ from $e^+e^-$ annihilations and their phenomenological implications are reviewed. Possible new sources on $N^*$ production from $e^+e^-$ annihilations are discussed.
\end{abstract}

\section{Introduction}
\label{intro}
Historically the study of spectroscopy at various microscopic levels of matter proves to be a powerful tool to explore the relevant structures and interactions. About a hundred years ago, the study of atomic spectroscopy revealed the quantum physical picture for atoms and played a important role for the development of quantum mechanics.  Around the middle of last century, the study of nuclear spectroscopy led to the two Nobel prize-winning works: nuclear shell model and collective motion model. With the quark model developed in the early 1960s, it became clear that the hadrons are not elementary particles, but composed of quarks and antiquarks. In the classical quark model, a baryon is composed of three quarks and a meson is composed of one quark and one antiquark. The only stable hadron is the proton which was proposed to be composed of two u-quarks and one d-quark. Since then, the protons were used as targets to be bombarded by pion, electron, photo beams to explore the spectroscopy of excited nucleons, $N^*$ resonances~\cite{Klempt:2009pi,Aznauryan:2011qj,Crede:2013sze,Patrignani:2016xqp}. 

With accumulation of half century on the $N^*$ spectroscopy, two outstanding problems appeared for the classical simple 3q constituent quark model. The first outstanding problem is that the mass-order for the lowest excited states is reversed. In the simple 3q constituent quark model, the lowest spatial excited baryon is expected to be a (uud)
$N^*$ state with one quark in orbital angular momentum $L=1$ state, and hence should have negative parity. Experimentally~\cite{Patrignani:2016xqp}, the lowest negative parity $N^*$ resonance is found to
be $N^*(1535)$, which is heavier than $N^*(1440)$ of positive parity. The second outstanding problem is
that in many of its forms the classical 3q quark model predicts a substantial number of
`missing $N^*$ states' around 2 GeV/$c^2$, which have not so far been observed~\cite{Capstick1}.

The first problem suggests that we should go beyond the simple 3q quenched quark model. It can be reconciled by taking these $N^*$s as meson-baryon dynamically generated
states~\cite{Oller:2000ma,Kaiser:1995eg,Oller:2000fj,Inoue:2001ip,GarciaRecio:2003ks,Hyodo:2002pk} or considering large 5-quark components in them~\cite{Zou:2007mk,Liu:2005pm,Helminen:2000jb}.

For the second problem, non-observation of these `missing $N^*$ states' does not necessarily mean that they do not
exist. Their couplings to $\pi N$ and $\gamma N$ may be too weak to be observed by
presently available $\pi N$ and $\gamma N$ experiments~\cite{Capstick1}. Other production processes should be
explored. Joining the effort on studying the excited nucleons, $N^*$ baryons, BES started a baryon resonance program~\cite{Zou1} at Beijing Electron-Positron Collider (BEPC) just before the start of new century. The $J/\psi$ and $\psi'$ experiments at BES provide an excellent place for studying excited nucleons and hyperons -- $N^*$, $\Lambda^*$, $\Sigma^*$ and $\Xi^*$ resonances~\cite{Zou2,Asner:2008nq}. 

In the following, the major experimental results on $N^*$ from $e^+e^-$ annihilations for last 20 years and some of their interesting phenomenological implications are reviewed.

\section{$N^*$ production from $\bar cc$ decays}
\label{sec:2}

Since 2001, BES/BESII/BESIII Collaborations have published their results on $N^*$ production from $J/\psi\to\bar pp\eta$~\cite{Bai:2001ua}, $p \bar{n}\pi^-+c.c.$~\cite{Ablikim:2004ug}, $p \bar{p}\pi^0$~\cite{Ablikim:2009iw}, $pK^-\bar\Lambda+c.c.$~\cite{Ablikim:2004dj}, $\bar nK^0_S\Lambda$~\cite{Ablikim:2007ec}, $\bar  pp\omega$~\cite{Ablikim:2007ac}, $\bar  pp\phi$~\cite{Ablikim:2015pkc}, $\bar pp\pi^0\eta$~\cite{Ablikim:2014dnh}, and $\psi(2S)\to\bar pp\eta$~\cite{Ablikim:2013vtm}, $p \bar{n}\pi^-+c.c.$~\cite{Ablikim:2006aha}, $p \bar{p}\pi^0$~\cite{Ablikim:2012zk}, $\bar{p} K^+ \Sigma^0$~\cite{Ablikim:2012ff}, and $\chi_{cJ}\to p\bar{n}\pi^{-}$~\cite{Ablikim:2012ih}, $p\bar{n}\pi^{-}\pi^{0}$~\cite{Ablikim:2012ih}, $\bar{p} K^+ \Lambda$~\cite{Ablikim:2012ff}, and $\psi(3770)\to p \bar{p}\pi^0$~\cite{Ablikim:2014kxa}. Some interesting insights on the $N^*$s have been gained through this novel source of information.

\begin{figure}
  \includegraphics[width=0.5\textwidth]{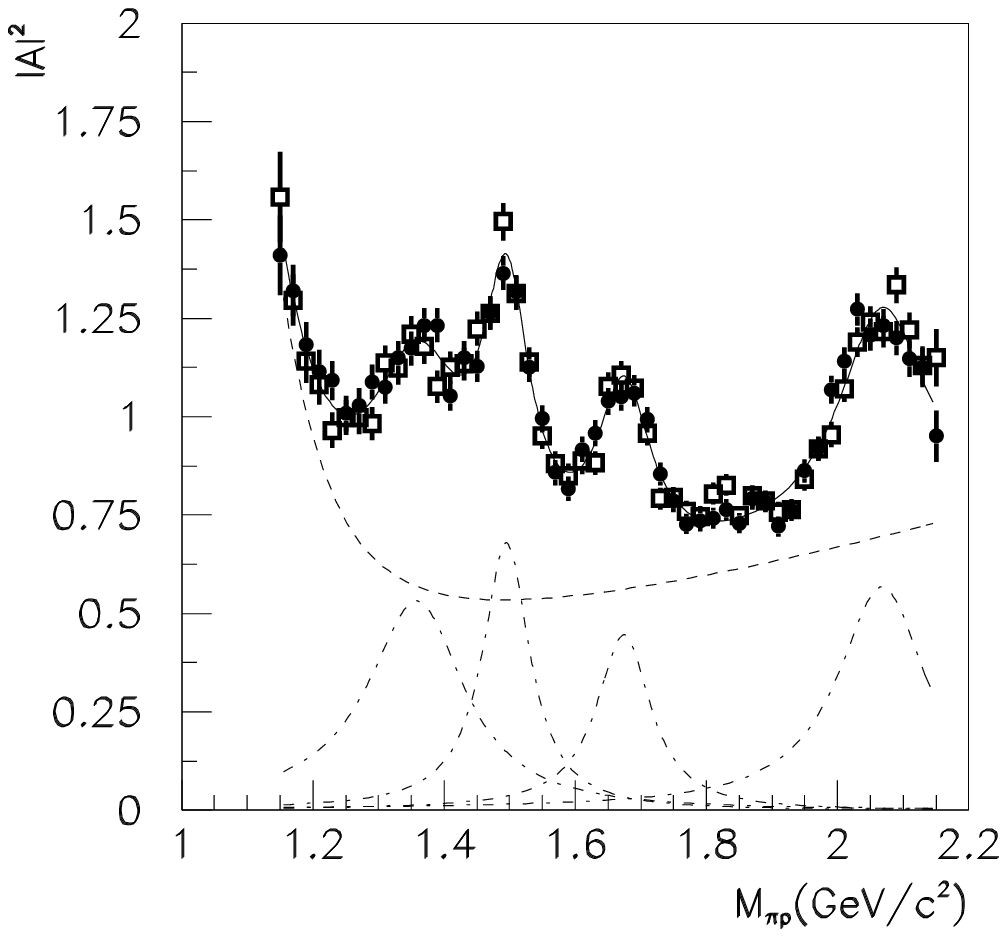}
  \includegraphics[width=0.45\textwidth]{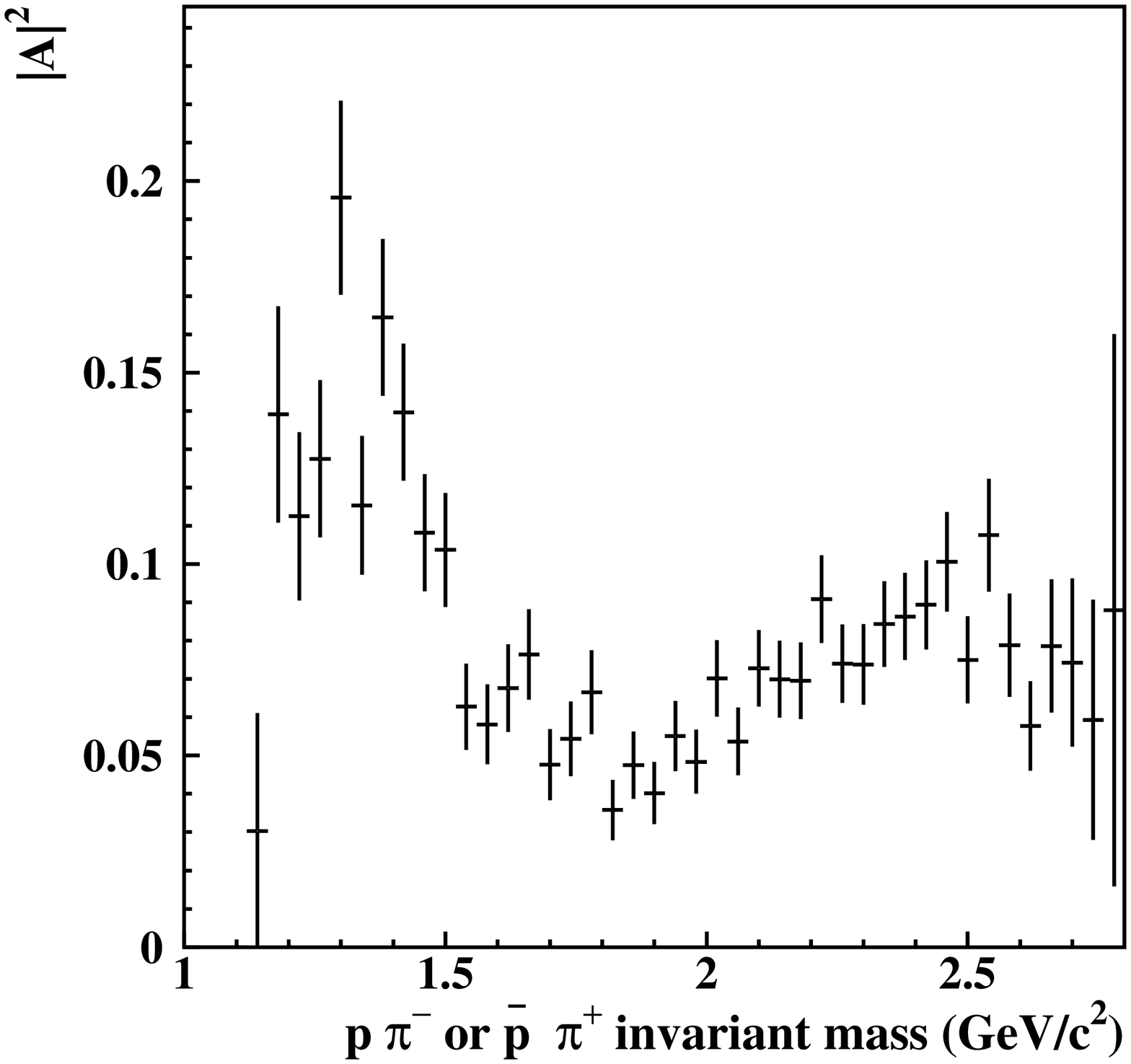}
\caption{\label{am} Data corrected by MC simulated efficiency and
phase space versus $p \pi^-$ (or $\bar{p} \pi^+$) invariant mass for $J/\psi \to p \bar{n}
\pi^-+c.c.$~\cite{Ablikim:2004ug} (left) and $\psi' \to p \bar{n}
\pi^-+c.c.$~\cite{Ablikim:2006aha} (right).}
\end{figure}

\begin{figure}
  \includegraphics[height=2.0in,width=2.1in]{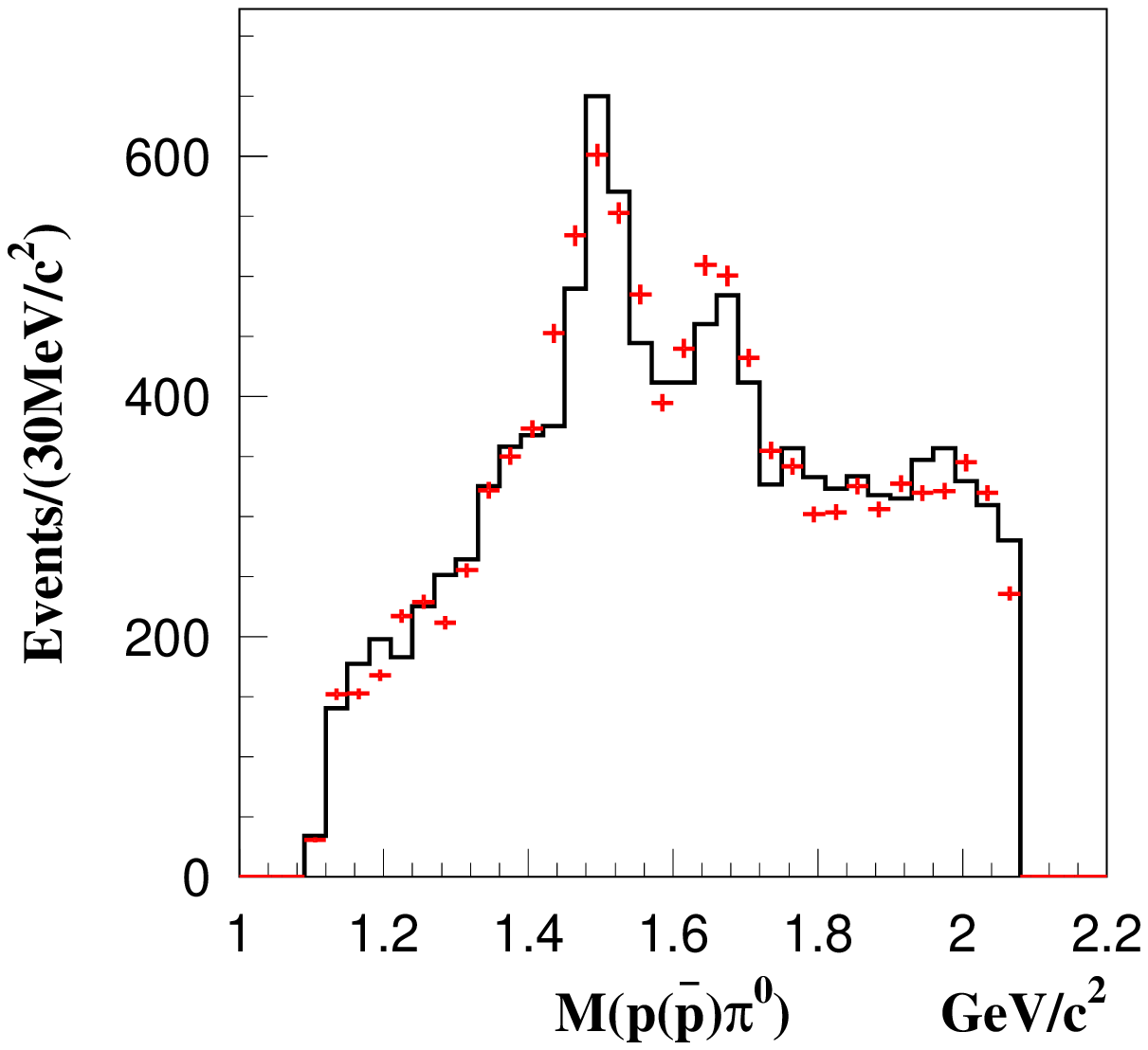}
  \includegraphics[height=2.25in,width=2.4in]{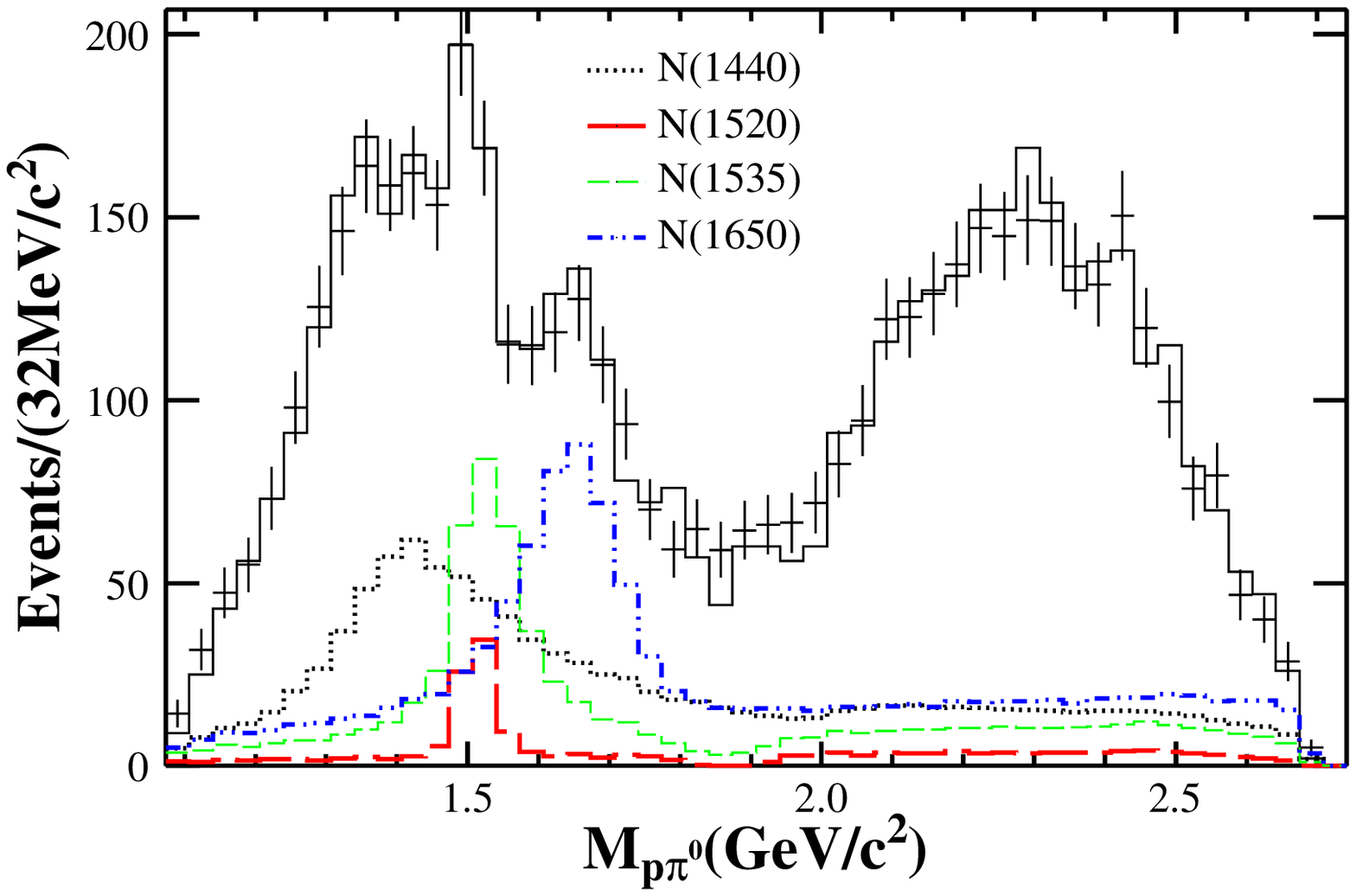}
\caption{\label{fig2} $p \pi^0$ (or $\bar{p} \pi^0$) invariant mass for $J/\psi \to p \bar{p}
\pi^0$~\cite{Ablikim:2009iw} (left) and $\psi' \to p \bar{p}
\pi^0$~\cite{Ablikim:2012zk} (right).}
\end{figure}

In Fig.\ref{am}, the invariant mass data corrected by MC simulated efficiency and
phase space versus $p \pi^-$ (or $\bar{p} \pi^+$) invariant mass for $J/\psi \to p \bar{n}
\pi^-+c.c.$ and $\psi' \to p \bar{n}\pi^-+c.c.$ are shown together for a comparison.
Similarly, in Fig.\ref{fig2}, $p \pi^0$ (or $\bar{p} \pi^0$) invariant mass for $J/\psi \to p \bar{p}
\pi^0$ and $\psi' \to p \bar{p}\pi^0$ are presented. Compared with $N\pi$ invariant mass spectrum from $\pi p$ or $\gamma p$ reactions, an obvious phenomena is that there are more $N^*$ peaks meanwhile without the strong $\Delta$ peak.
This is because $\psi$ annihilates into a baryon-antibaryon pair through three gluons and conserves isospin.
The $N\pi$ recoiling an anti-proton is limited to be isospin 1/2. So the charmonium annihilation provides a nice isospin filter. Due to the non-presence of the strong $\Delta$ peak in other reactions, the $N^(1440$ peak was observed for the first time directly from $\pi N$ invariant mass spectrum. Besides several well known $N^*$ resonances around 1520 MeV and 1670 MeV, three new $N^*$ resonances above 2 GeV were found through delicate partial wave analyses. They are $N^*(2040) 3/2^+$, $N^*(2300) 1/2^+$ and $N^*(2570) 5/2^-$. An additional advantage of this reaction is that it not only selects isospin 1/2 states but also suppresses high spin states due to the short range interaction involved in the $\bar cc$ annihilation that generates the $N\pi$ system. The suppression of higher spin states greatly simplifies partial wave analysis.

Another interesting phenomena is that the $N^*(1440)$ is produced much stronger from $\psi(2S)$ than from $J/\psi$.
There are two common features for $\psi(2S)$ and $N^*(1440)$: 1) they are supposed to be “radial” excitation of $J/\psi$ and nucleon, respectively, in the simple quenched quark model; 2) they were found experimentally to have large coupling to $\sigma J/\psi$ and $\sigma N$, respectively. In unquenched quark models, “radial” excitations like to pull out $\bar q^2q^2(0+)$ from sea, hence favor transition between each other. This unquenched picture not only gives a natural explanation of much enhanced $N^*(1440)$ production from $\psi(2S)$ than $J/\psi$, may also explain the long-standing $\rho\pi$ puzzle~\cite{Asner:2008nq} from $\psi(2S)$ and $J/\psi$ decays, {\sl i.e.}, $\psi(2S)$ tends to decay into $\rho(2S)\pi$ while $J/\psi$ tends to decay into $\rho\pi$. CLEO Collaboration also studied $\psi(2S)\to\bar pp\pi^0$ channel and got a similar strong $N^*(1440)$ peak~\cite{Alexander:2010vd}. There is no obvious $N^*(1440)$ peak for $e^+e^- \to p \bar p \pi^0$ in the vicinity of the $\psi (3770)$~\cite{Ablikim:2014kxa}.

\begin{figure}
\includegraphics[scale=0.19]{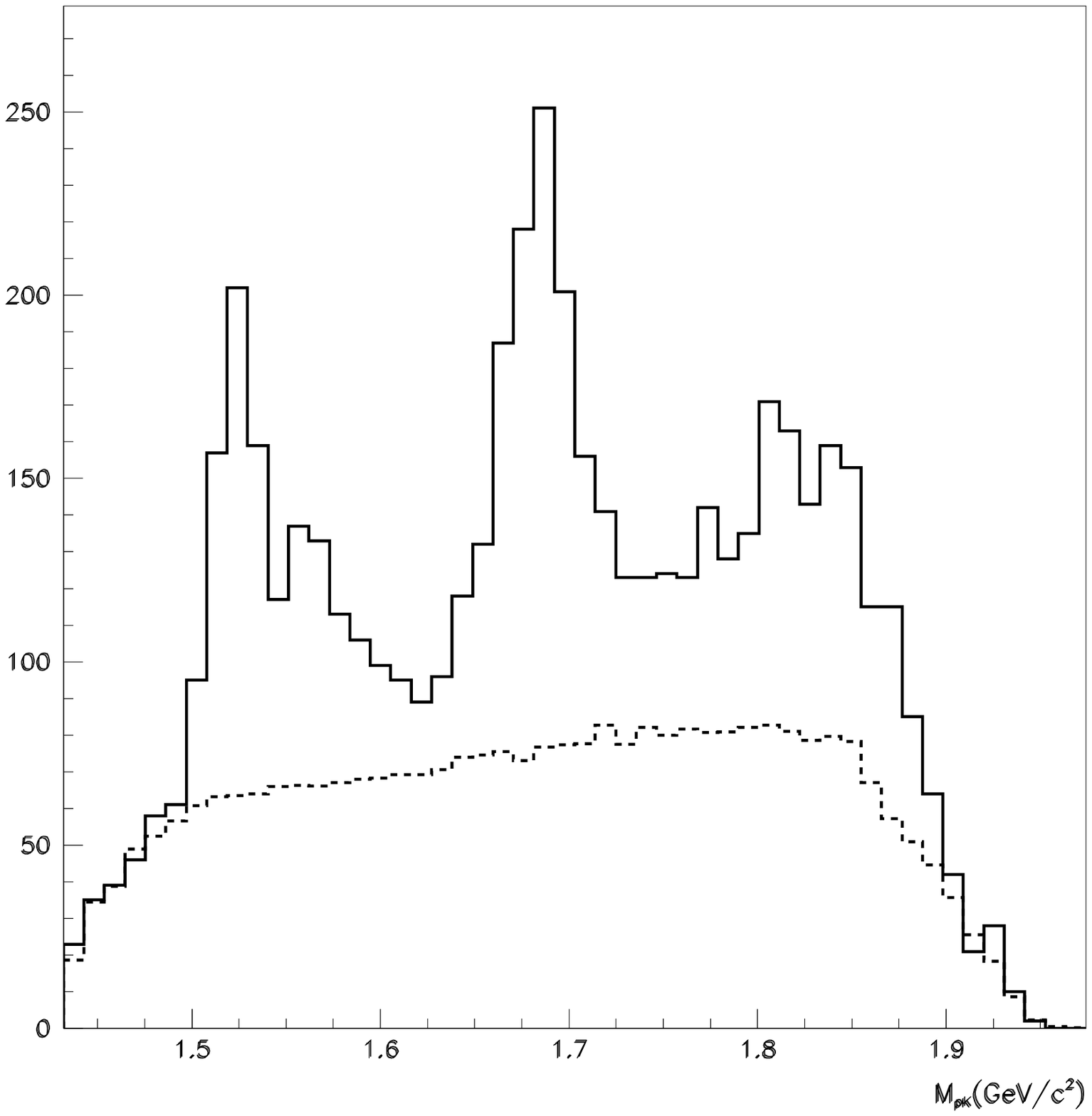}
\includegraphics[scale=0.19]{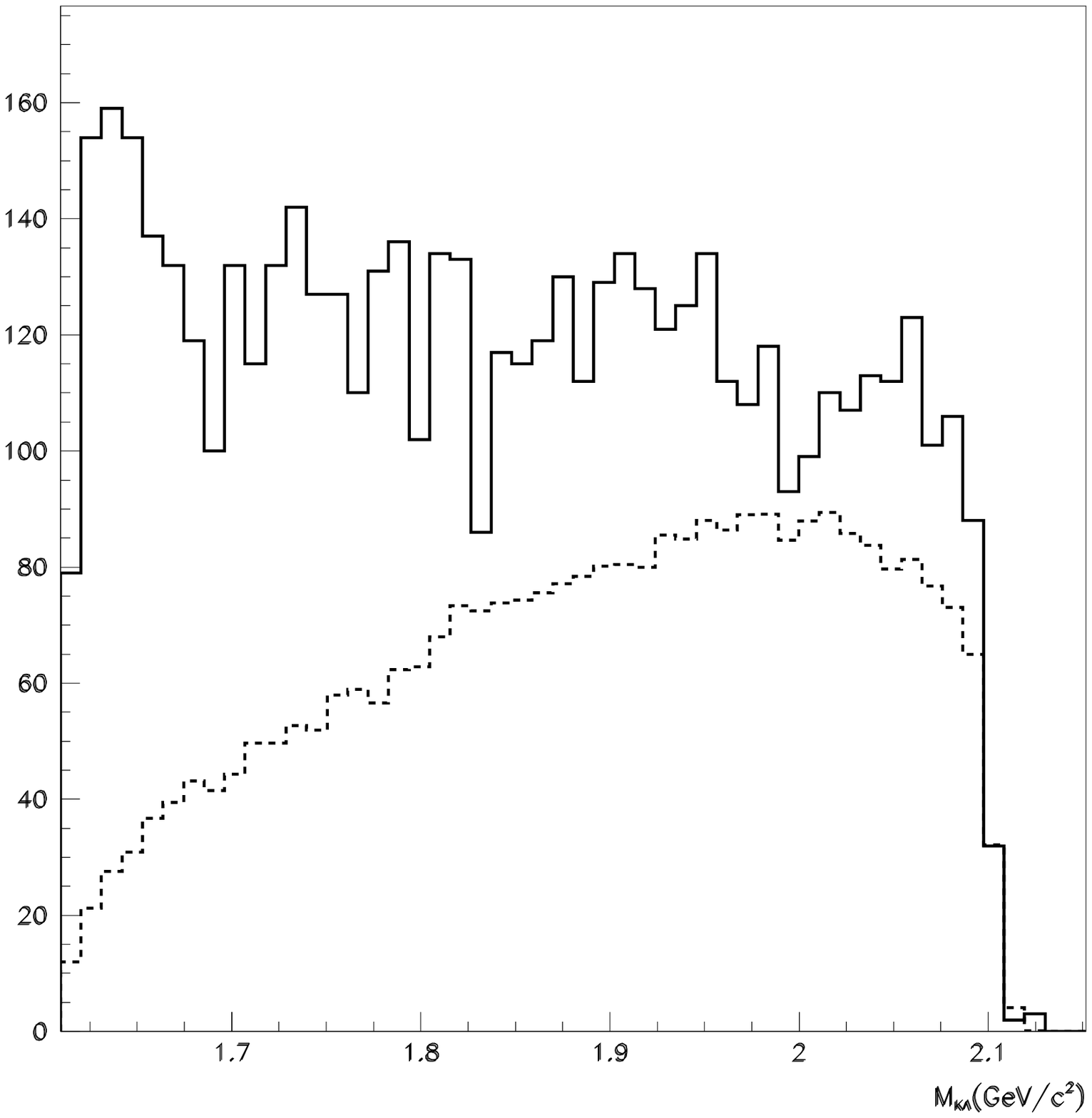}
\includegraphics[scale=0.19]{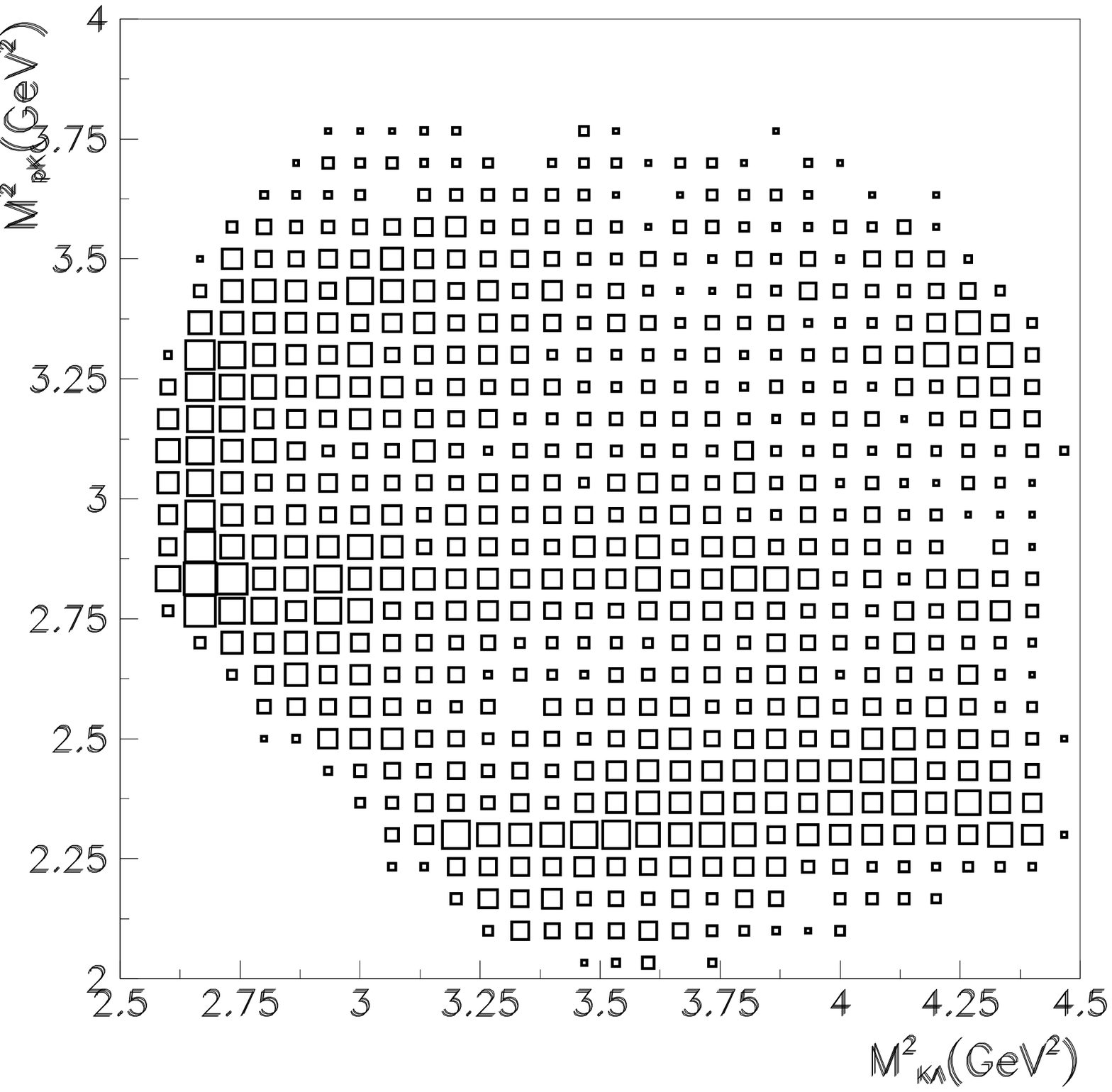}
\caption{\label{fig3} $pK$ (left) and $K\Lambda$
(middle) invariant mass spectra for $J/\psi\to
pK^-\bar\Lambda$+c.c., compared with phase space distribution;
right: Dalitz plot for $J/\psi\to pK^-\bar\Lambda$+c.c.~\cite{Ablikim:2004dj}}
\end{figure}

In Fig.\ref{fig3}, the Dalitz plot and corresponding invariant mass spectra are presented for $J/\psi\to pK^-\bar\Lambda$ and $\bar pK^+\Lambda$ channels~\cite{Ablikim:2004dj}. There are clear $\Lambda^*$ peaks at 1.52 GeV, 1.69 GeV and 1.8 GeV in $pK$ invariant mass spectrum, and $N^*$ peaks
near $K\Lambda$ threshold, 1.9 GeV  and 2.05 GeV for $K\Lambda$
invariant mass spectrum. The $N^*$ peak near $K\Lambda$ threshold is most probably due to $N^*(1535)$. Combined with information on $N^*(1535)$ from $J/\psi\to\bar pp\eta$~\cite{Bai:2001ua} as well as COSY data on $pp\to pK^+\Lambda$, a large coupling to $K\Lambda$ was found for the $N^*(1535)$~\cite{Liu:2005pm}. This supports it to be a $K\Sigma$-$K\Lambda$ dynamically generated state with large hidden strangeness component. Extending this picture from strangeness to charm and beauty, super-heavy $N^*$ with hidden charm~\cite{Wu:2010jy} or hidden beauty~\cite{Wu:2010rv} were predicted to exist around 4.3 GeV and 11 GeV, respectively. Two super-heavy $N^*$ states with hidden charm were later discovered by LHCb experiment~\cite{Aaij:2015tga} from $\Lambda_b$ decays. Their meson partners $Z_c$ states were also discovered by BESIII Collaboration~\cite{Ablikim:2013mio,Ablikim:2013wzq} and other experiments as reviewed in Refs.\cite{Chen:2016qju,Guo:2017jvc}.

%
%

\section{Hyperon production and Prospects}
\label{sec:3}

Besides $N^*$ resonances, some hyperon resonances were also studied by BESIII from $J/\psi\to\gamma\Lambda\bar\Lambda$~\cite{Ablikim:2012bw}, and $\psi(2S)\to\bar{p} K^+ \Sigma^0$~\cite{Ablikim:2012ff}, $\Lambda\bar\Sigma^{\pm}\pi^{\mp}+c.c.$~\cite{Ablikim:2013xra}, $\psi(2S)\to K^{-} \Lambda \bar{\Xi}^{+} +c.c.$~\cite{Ablikim:2015apm}.

\begin{figure}
  \includegraphics[height=2.0in,width=2.0in]{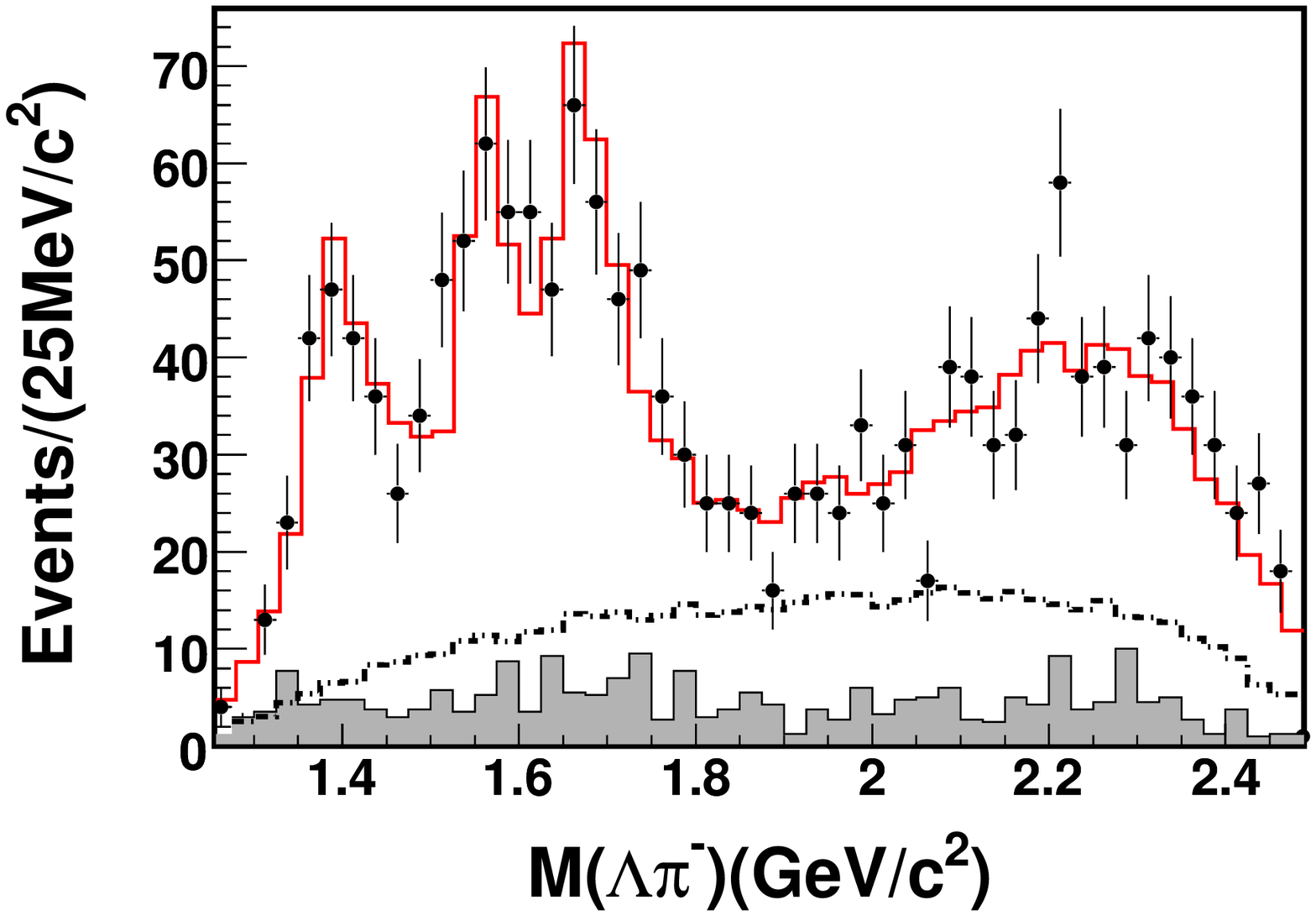}
  \includegraphics[height=2.1in,width=2.1in]{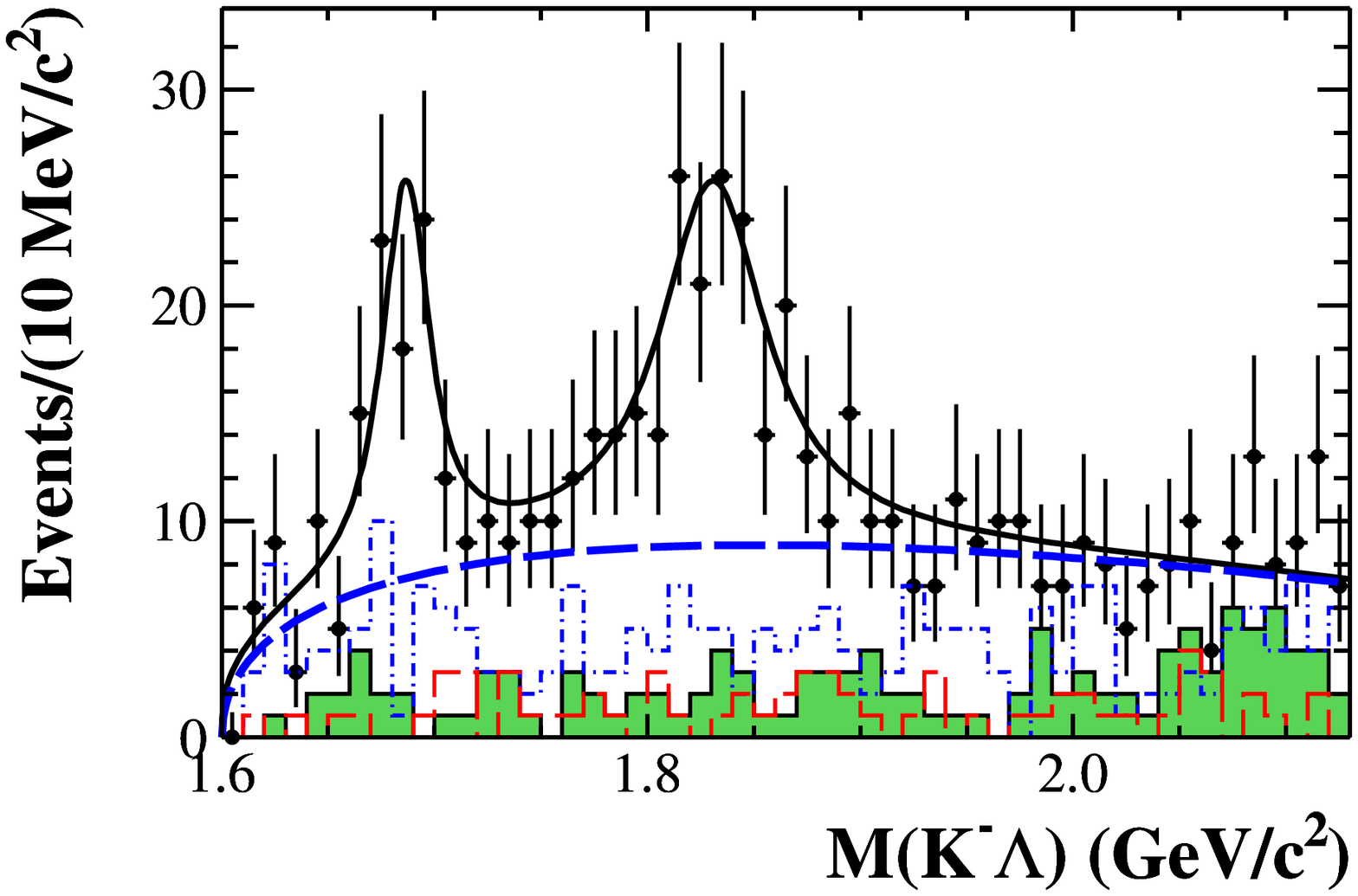}
\caption{\label{fig4} $\Lambda\pi^-$ invariant mass for $\psi(2S) \to \Lambda\bar\Sigma^{\pm}\pi^{\mp}$~\cite{Ablikim:2013xra} (left) and $K^-\Lambda$ invariant mass for $\psi(2S)\to K^{-} \Lambda \bar{\Xi}^{+} +c.c.$~\cite{Ablikim:2015apm} (right).}
\end{figure}

Two typical invariant mass plots for hyperon resonances are shown in Fig.\ref{fig4}. Clear resonance peaks are observed for $\Sigma^*$ and $\Xi^*$ resonances. There is a clear $\Sigma^*$ peak around 1580 MeV which can be fitted well with the 1-star $\Sigma(1580) 3/2^-$ resonance of PDG~\cite{Patrignani:2016xqp}. In 2012, by analyzing $K^-p\to\pi^0\Lambda$ data, we also found some evidence for a $\Sigma^*(3/2^-)$ resonance around 1542 MeV~\cite{Gao:2012zh}. A $\Sigma^*(3/2^-)$ around 1560 MeV was expected by unquenched quark model~\cite{Helminen:2000jb}. 

For $e^+e^-$ annihilations at energies above $\Lambda_c\bar\Lambda_c$ threshold, the $\Lambda_c$ decays provide a new source on the $N^*$ and hyperon spectroscopy.  Recently, Belle Collaboration observed a very narrow $\Lambda^*$ peak around 1670 MeV in the $pK$ invariant mass spectrum in $\Lambda^{+}_{c} \to p K^{-} \pi^{+}$~\cite{Yang:2015ytm}. This is consistent with a previous observation of a very narrow $\Lambda^*(1670) 1/2^-$ from analyzing $K^-p\to\Lambda\eta$ data~\cite{Liu:2012ge}. If it is confirmed, it would be a natural candidate of $[ud]ss\bar s$ pentaquark state which can only decay to $\Lambda\eta$ through strongly suppressed D-wave decay. It is important to check its existence through $\Lambda\eta$ invariant mass spectrum of $\Lambda^{+}_{c} \to \Lambda\eta\pi^+$. For $e^+e^-$ annihilations at energies above $\Lambda_b\bar\Lambda_b$ threshold at super-B or super-Z factories, its $\Lambda_b$ decays would provide a much cleaner source than LHCb experiment to look for super-heavy $N^*$ and hyperon resonances with hidden-charm.

With further accumulation on charmonium decays, there are many more interesting channels can be explored, such as $\bar\Omega\Xi\bar K$, $\bar\Xi\Xi\pi$, $\bar\Lambda\Lambda\gamma$, $\bar\Sigma\Lambda\gamma$, $\bar\Sigma\Sigma\gamma$, $\bar\Xi\Xi\gamma$, etc., with $\Omega\to\Lambda K^-$ and $\Xi\to\Lambda\pi$. While CEBAF at JLab has advantage for studying radiative decays of $N^*$ and $\Delta^*$, BESIII may have advantage to study radiative decays of $\Lambda^*$, $\Sigma^*$ and $\Xi^*$.   To complete $N^*$, $\Lambda^*$, $\Sigma^*$, $\Xi^*$ spectra and establish the lowest $\Lambda^*$, $\Sigma^*$, $\Xi^*$ and $\Omega^*$ with partial wave analysis, a super $\tau$-charm factory may be needed.




\end{document}